\documentstyle[12pt,dina4p,slash,preprint]{article}

\begin{document}
\def\undersim#1{\mathord{\vtop{\ialign{##\crcr
$\hfil\displaystyle{#1}\hfil$\crcr\noalign{\kern1.5pt\nointerlineskip}
$\hfil\sim{}\hfil$\crcr\noalign{\kern1.5pt}}}}}
\def\mayeq{\undersim{>}}
\def\meneq{\undersim{<}}
\preprint{FTUAM 38/93}
\date{November, 1993}
\title{On the Evaluation of Threshold Effects in Processes Involving
Heavy Quarks \footnote{Research partially supported by the U.S.
Department of Energy and CICYT, Spain}}
\author{F.J. Yndur\'ain\\Departamento de F\'{\i}sica Te\'orica,
C-XI,\\Universidad
Aut\'onoma de Madrid, Cantoblanco\\E-28049, Madrid, Spain}
\maketitle
\begin{abstract}
A detailed evaluation is presented of production of heavy fermions
(particularly $t\overline{t}$ quarks) by a vector current, in the region
around threshold. This includes bound states as well as the region above
threshold, to the same degree of accuracy; in both cases radiative and
nonperturbative corrections are included. The contribution of this
region to the vacuum polarization (``threshold effects'') is calculated
and compared with a perturbative calculation; the importance of the
threshold effects is found substantially smaller than claimed by some
authors.

Open fermion-antifermion production is also discussed.
\end{abstract}

\newpage

\section{Introduction}
In this note we consider the structure of the threshold region for
$q\overline{q}$ production (also for lepton production) and, in greater
detail, its influence on certain quantities involving these quarks. With
respect to the last, we will study the photon propagator (or the vector
part of a $Z$ propagator), $\Pi (s)$, for a momentum squared $s$;
evaluating in particular the influence of a loop of heavy fermions
(i.e., with masses such that $4m^2 \gg s$) on this propagator. We will
focus our interest on the influence on $\Pi (s)$ of the area around the
fermion threshold $s=4m^2$.

The motivation for our work is the recent appearance of a number of
papers dealing with such effects, in some cases incorrectly (in our
opinion). To be precise we will refer to the work of Kniehl and
Sirlin$^{[1]}$ as the most recent and complete of them, but similar
criticism applies to much of the literature on the subject. To clarify
matters we will start by considering the QED case (with lepton loops)
where several of the problems can be understood easiest, moving
afterwards to the more interesting case of QCD (in particular involving
the $t$ quark loop and threshold).

We establish our notation. The current associated to a fermion with
field $\psi$ and mass $m$ will be written
\begin{equation}
J_{\mu} (x) = \overline{\psi} (x) \gamma_{\mu} \psi (x);
\end{equation}
in the QCD case a trace over omitted colour indices is understood. We
define the two-point tensor $\Pi^{\mu \nu}$ and function $\Pi$ by
\begin{equation}
\Pi_{\mu \nu} (p) = i \int d^4 x e^{ip \cdot x} < T J_{\mu} (x) J_{\nu}
(0)> = (p^2g_{\mu \nu} - p_{\mu} p_{\nu}) \Pi (p^2).
\end{equation}

The function $\Pi (s)$ can be proved, quite independently of
perturbation theory, to satisfy a dispersion relation: for $0 < s <
4m^2$,
\begin{equation}
\Pi (s) - \Pi (0) = \frac{s}{\pi} \int^{\infty}_{4m^2} ds' \frac{Im \Pi
(s')}{s' (s'-s)}.
\end{equation}

Of course by expanding both sides it follows that (3) is also valid
order by order in perturbation theory. The key point in our discussion
is precisely this: because (3) is really an identity (being just the
representation of $\Pi (s)$ through a Cauchy integral), nothing can be
learned on the l.h.s. of (3) from perturbative calculation of the r.h.s.
there. New knowledge will emerge only if independent input {\em of a
better quality is used}: for example, very precise experimental data, or
data where the theoretical evaluations are not valid, or exact,
nonperturbatuve effects {\em larger} than the perturbatively known ones.

\section{QED}

Here we consider that the particles described by $\psi$ are leptons. To
one loop we expand $\Pi (s) - \Pi (0)$ in powers of $s/m^2$, assumed to
be small:
\begin{equation}
\Pi^{(1)} (s) - \Pi^{(1)} (0) = \frac{1}{4\pi^2} \frac{s}{15 m^2}
\left\{ 1 + \frac{3}{28} \frac{s}{m^2}+ \ldots \right\}.
\end{equation}
For the higher order corrections only the leading term in $s/m^2$ will
be taken into account. To two loops, the correction to (4) is known
since a long time (ref. [2]; a more easily accesible reference is [3]).
We have, to terms $0(s^2/m^4)$, $0(\alpha s/m^2)$,
\begin{equation}
\Pi^{(1+2)} (s) - \Pi^{(1+2)} (0) = \frac{1}{4\pi^2} \frac{s}{15m^2}
\left\{ 1 + \frac{3}{28} \frac{s}{m^2} + \frac{205}{54}
\frac{\alpha}{\pi} \right\}.
\end{equation}
For the {\em three loop} calculation we will consider, with a view to
QCD, the following situation: besides the lepton described by $\psi$
there is another one with a mass $\mu < m$. (For example, we could have
$\psi$ represent a muon and the lighter fermion be an electron.) The
renormalization group may then be used to write
\begin{eqnarray}
\Pi^{(1+2+3)} (s) - \Pi^{(1+2+3)} (s) = \frac{1}{4\pi^2} \frac{s}{15m^2}
\nonumber \\
\times \left\{ 1+\frac{3}{28} \frac{s}{m^2}+ \frac{205}{54}
\frac{\alpha}{\pi} + \frac{205}{162} \left( \frac{\alpha}{\pi}
\right) ^2 \left(
\log \frac{m^2}{\mu^2} +a \right) \right\}, \ \ s \meneq \mu ,
\end{eqnarray}
and the constant $a$ can only be obtained with the full three loop
calculation.

In ref. [1], the authors propose to improve an estimate like that in (5)
by being more accurate in the treatment of the threshold region, $s
\simeq 4m^2$, using then a dispersion relation. For the imaginary part
of $\Pi$ they take
\subequationsa
\begin{eqnarray}
Im \Pi (s) & = & \sum_{n} \frac{1}{M_n} | R_{n0} (0) |^2 \delta
(s-M^2_n) \nonumber \\
		  & + & \theta (s - 4m^2) \frac{v (3-v^2)}{24\pi} \left[
1+\frac{\pi \alpha}{2v} + \ldots \right], \ \ s \simeq 4m^2.
\end{eqnarray}
The first term in the r.h.s. is the contribution of the bound states
with mass $M_n = 2m (1-\alpha^2/8n^2)$ and radial wave function
\begin{equation}
| R_{n0} (0)|^2 = \frac{m^3\alpha^3}{2n^3}.
\end{equation}
\endsubequationsa
$v=(1-4m^2/s)^{1/2}$ is the velocity of the leptons in the c.m. $n$ is
the principal quantum number.

Eq. (7) is {\em inconsistent}. For small $v$ higher orders in
perturbation theory give increasing corrections, just above threshold.
These may be summed explicitely to the factor
\begin{equation}
F(v) = \frac{\pi \alpha /v}{1- e^{-\pi \alpha /v}},
\end{equation}
the square of the wave function for lepton-antilepton Coulomb scattering
at the origin\footnote{This was first noticed by Fermi (ref. 4) when
studying Coulomb corrections to $\beta$ decay; $F$ is at times known as
the Fermi factor, and the effect as the Fermi-Watson final state
interaction theorem}. So we should replace (7) by
\begin{eqnarray}
Im\Pi_{thr} (s) & = & \sum_n \frac{1}{M_n} |R_{n0}(0)|^2 \delta
(s-M^2_n) \nonumber \\
		& + & \theta (s-4m^2) \frac{v (3-v^2)}{24 \pi} \frac{\pi
\alpha /v}{1- exp (-\pi \alpha /v)}.
\end{eqnarray}
The reason why (7) is inconsistent, and (9) is consistent, is easily
understood recalling the following facts. The bound states are a
nonperturbative phenomenon, so you should also use the nonperturbative
form of $Im \Pi$ above threshold. Indeed, if we continue $F(v)$
analytically for $s < 4m^2$, $v$ goes over $i|v|$ and one finds that $F
(i|v|)$ has poles at the precise values $s = M^2_n$: the approximations
of including poles and $F(v)$ are at the same level.

The full novel effect of the nonperturbative contributions around
threshold is then obtained subtracting from (9) the purely perturbative
expression. To two loops,
\begin{equation}
Im \Pi^{(1+2)} (s) = \theta (s-4m^2) \frac{v(3-v^2)}{24\pi} \left[ 1+
\frac{\pi \alpha}{2v} - \frac{4 \alpha}{\pi} + \frac{\pi v \alpha}{3}
\ldots \right] .
\end{equation}
Because (9) is only valid above threshold for $v \ll \pi \alpha$, a
reasonable estimate is obtained by cutting the dispersive integral at an
$s_0$ such that $v= \frac{\pi}{2} \alpha $. Thus we get the threshold
effect given as
\begin{equation}
\Delta_{thr} (s)=\frac{s}{\pi} \int^{s_0}_{4m^2} ds' \frac{Im \Pi_{thr}
(s) - Im \Pi^{(1+2)} (s)}{s' (s'-s)} .
\end{equation}
The integral in (11) can be made easily; to relative errors $\alpha^2$,
$s/m^2$, $1/e^{\pi}$ we get
\begin{equation}
\Delta_{thr} (s) = \frac{1}{4\pi^2} \frac{s}{15m^2} \frac{15\pi (\zeta
(3) + \pi^2/12) \alpha^3}{16} .
\end{equation}
The term in $\zeta (3)$ comes from integrating the poles, and the rest
from the region above threshold. (12) is to be compared with the
perturbative piece (6), i.e., the full quantity would be
\begin{equation}
\Pi^{(1+2+3)} + \Delta_{thr} .
\end{equation}
It is however apparent that, in spite of its nonperturbative origin,
$\Delta_{thr}$ is of order $\alpha^3$. To be able to add it meaningfully
to a perturbative evaluation it would be necessary that the last be of
the same precision; i.e., one would require, at the very least, the
complete evaluation of the {\em three loop} contribution and, to be
rigorous, the {\em four loop} one. For example, one has, numerically,
\[ \frac{15 \pi (\zeta (3) + \pi^2/12)}{16} \alpha^3 = 2.32 \times
10^{-6} \]
while the three loop piece of (6) is (with $m=m_{\mu}$, $\mu =m_e$)
\[ \frac{205 \log m^2_{\mu}/m^2_e}{162\pi^2} \alpha^2 = 7.239 \times
10^{-5} \]
i.e., the three-loop term (which is not fully known!) is expected to be
almost two orders of magnitude larger than the threshold effects.

\section{QCD}

There are two cases of interest in QCD: that of open or bound
$q\overline{q}$ production close to threshold, and the influence of this
when $q$ is a $t$ quark, and $s=M^2_z$. We will discuss both cases. When
making numerical evaluations involving the $t$ quark we will assume
\[ m_t^2 / M^2_z =3 \ ; \ \ \alpha_s (M^2_z) = 0.115 \pm 0.01.	\]

Before entering the main discussion a few words have to be said on the
influence of nonperturbative effects on the threshold properties of
heavy quarks, specifically toponium, for the litterature is riddled with
misunderstandings (as e.g. is the case in refs. 5, 6, besides ref. 1).
To be precise, these authors assume that the $q\overline{q}$ potential
for $t$ quarks will incorporate, in addition to a Coulombic piece,
\[ - C_F \alpha_s /r, \]
a ``confining'' linear potential
\begin{equation}
Kr, \ K^{1/2} \sim 0.5 \ GeV.
\end{equation}

Now, and as should be well known, the confining potential has a linear
form only at long distances; that is to say, (14) only holds for $r \gg
\Lambda^{-1}$ with $\Lambda \sim 0.2  \ GeV$. Since the radius of
$t\overline{t}$ is of the order of 1/20 GeV, it should be clear that
what is relevant for toponium is not the long range behaviour of the
confinement forces, but their {\em short distance} structure. In this
context Leutwyler$^{[7]}$ has shown a long time ago that the {\em short
distance} effects of confinement make themselves manifest through the
contributions of the gluon condensate $< \alpha_s G^2>$ to the
properties of heavy quark bound states; a fact that also emerges from
the classical papers of SVZ$^{[8]}$ (This last we will discuss later).
What one has from Leutwyler's analysis is that, at short distances, the
confinement forces may be simulated by a potential behaving like
\begin{equation}
\frac{\pi \epsilon_{10} <\alpha_s G^2>}{60 C_F \alpha_s} r^3, \ \ r \ll
\Lambda^{-1},
\end{equation}
where $\epsilon_{10} = 1.468$ and the gluon condensate is
\[ < \alpha_s G^2> \simeq 0.042 \ GeV^4 . \]
For $c\overline{c}$, (14) contributes dominantly\footnote{For the
$c\overline{c}, b\overline{b}$ analysis see ref. 9. Note that we do not
mean that one should forget (4); a linear potential should be used to
describe states with $n \mayeq 1$ for $c\overline{c}$, $n \mayeq 2$ for
$b\overline{b}$ and $n \mayeq 5$ for $t\overline{t}$.} already at $n=1$.
For $b\overline{b}$, the contribution of (14) is small for $n=1$, and
large (for some quantities dominant) for $n=2$. For $t\overline{t}$ the
influence of (14) will be negligible up to $n \sim 5$: because of the
extremely short distance at which $t\overline{t}$ orbit each other,
toponium
should be very well described (for $n < 5$) by a Coulombic potential.
Contrarywise, a calculation of $n<5$ toponium states with a linear
potential will give highly misleading results. For example, (14) and
(15) are equal for $r_c \sim 1/0.33 \ GeV$. For distances $r \ll r_c$,
(14) gives oversized contributions; in particular, for $n \leq 3$, an
overestimate of nonperturbative effects by {\em two or more} orders of
magnitude.

Because a Coulombic approximation is good for toponium bound states up
to $n \sim 5$, and the sum over bound states converges very rapidly,
like $n^{-3}$, (cf. eqs. (7)) we are justified in treating them as
Coulombic all the way. We may thus take over our analysis for the QED
case with obvious modifications. We find,
\begin{eqnarray}
& {\displaystyle \Pi^{(1+2+3)} (s) = \frac{N_c}{4\pi^2} \frac{s}{15
m^2}} \nonumber
\\ & {\displaystyle \times \left\{ 1 - \frac{3}{28} \frac{s}{m^2} +
\frac{205}{54} \frac{C_F
\alpha_s (\mu^2)}{\pi} \left[ 1+\frac{\beta_0 (\log \mu^2 /m^2) \alpha_s
(\mu^2)}{4\pi} \right] - \frac{2\pi <\alpha_s G^2>}{21 m^4} \right\} .}
\end{eqnarray}
Here $N_c=3$, $C_F=4/3$, $\beta_0=11-2n_f/3$ with $n_f=5$ for toponium.
As in eq. (6), only the renormalization-group generated piece of the
three loop contribution viz., the term
\[ \frac{205}{54} \frac{C_F \alpha_s}{\pi} \frac{\beta_0 (\log
\mu^2/m^2) \alpha_s}{4\pi}, \]
is included. The nonperturbative piece in (16),
\[ - \frac{2\pi <\alpha_s G^2>}{21 m^4} , \]
is taken form the standard SVZ calculation$^{[8]}$.

For the threshold effects we write
\subequationsa
\begin{equation}
Im \Pi_{pole} (s) = N_c \sum_n \left[ 1+\frac{3\beta_0 \alpha_s}{2\pi}
\left( \log \frac{n\mu}{mC_F \alpha_s} + \psi (n+1)-1 \right) \right]
\frac{|\tilde{R}^{(0)}_{n0} (0)|^2}{M_n} \delta (s-M^2_n),
\end{equation}
\begin{equation}
|\tilde{R}^{(0)}_{n0} (0)|^2=C^3_F m^3 \tilde{\alpha}_s (\mu^2)^3/2n^3;
\ \tilde{\alpha}_s(\mu^2) \equiv \left[ 1+\left( \frac{11C_A-4T_F
n_f}{6} \gamma_E + \frac{31C_A-20T_F n_f}{36} \right)
\frac{\alpha_s}{\pi} \right] \alpha_s .
\end{equation}
\endsubequationsa
We have included the radiative corrections to the Coulombic wave
functions (the term in square brackets in (17)), which are
sizeable\footnote{In ref. 1 the radiative corrections are incorrectly
given by a formula like (17) but with the term in square brackets
replaced by $1+4C_F \alpha_s/\pi$; cf. eq. (4.2) of ref. 1. This is
manifestly wrong, as it fails to make the whole term renormalization
group invariant.}. These are obtained using the one loop
corrections$^{[10, 9]}$ to the QCD potential; the analytic expression we
use was taken from ref. 9.

The importance of confinement effects in (17) may be estimated by
considering for example the corrections to the ground state wave
function$^{[7]}$:
\begin{eqnarray}
|R_{10} (0)|^2 & \rightarrow & |R_{10} (0)|^2 \left\{ 1+\frac{53424\pi <
\alpha_s G^2>}{3825 m^4 C^6_F \alpha_s^6} \right\} \nonumber \\
	       & = & |R_{10} (0)|^2 \{ 1+ 3.0 \times 10^{-4} \} ,
\nonumber
\end{eqnarray}
as was announced, a minute alteration that we will neglect.

Above threshold, but for $v \sim 0$, we have the perturbative expression
\begin{equation}
Im \Pi^{(1+2)} (s) = \frac{N_c v}{8\pi} \left[ 1-\frac{v^2}{3} +
\frac{1}{2} \frac{\pi C_F \alpha_s}{v} + \ldots - \frac{\pi <\alpha_s
G^2>}{192 m^4 v^6} \right],
\end{equation}
and the leading nonperturbative contribution (the last term) may be read
of directly form ref. 8. These terms also contain singularities in
$\delta (v)$, $\delta '(v)$ (which we have not written in (18)) in such
a way that
\begin{eqnarray}
 & {\displaystyle - \frac{N_c}{4\pi^2} \frac{s}{15m^2} \frac{2\pi
<\alpha_s G^2>}{21m^4}} \nonumber \\
= & {\displaystyle \frac{sN_c}{8\pi^2} \int ds' v(s') s'^{-2} \left[ -
\frac{\pi
<\alpha_s G^2>}{192 m^4v(s')^6} + {\mbox singularities} \right].}
\nonumber
\end{eqnarray}

The reader interested in the details should consult ref. 8. Again, and
except for exceedingly small values of $v$, eq. (18) shows that one can
entirely neglect the confinement effects for the $t\overline{t}$ system.

For small values of $v$ higher orders in $\alpha_s/v$ may be summed just
as in the QED case. We find
\begin{eqnarray}
Im \Pi_{exact} (s) \simeq \frac{N_cv}{8\pi}
\left( 1-\frac{v^2}{3} \right) \frac{\pi C_F \tilde{\alpha}_s
(\mu^2)}{v} \left[ 1+ \left( \log \frac{\mu C_F \alpha_s}{4mv^2} -1
\right) \frac{\beta_0 \alpha_s}{2\pi} \right] \nonumber \\
\times \left\{ 1- exp \left[ \frac{-\pi C_F \tilde{\alpha}_s (\mu^2)}{v}
\left( 1+ \frac{\beta_0 \alpha_s}{2\pi} \left( \log \frac{\mu C_F
\alpha_s}{4mv^2} -1 \right) \right) \right] \right\} ^{-1},
\end{eqnarray}
$\tilde{\alpha}_s$ as in (17b). The terms containing $\beta_0$, as well
as the replacement of $\tilde{\alpha}_s$ in lieu of $\alpha_s$, embody
the radiative corrections to the wave function at the origin evaluated
(as for bound states) with the help of the $q\overline{q}$ potential,
including one-loop radiative corrections.$^{[9, 10]}$ It should be noted
that these {\em diverge} as $v \rightarrow 0$: confinement results from
a collaboration of perturbative and nonperturbative effects. Putting in
numbers one may check that, for $c\overline{c}$ and (to a lesser extent)
$b\overline{b}$ the nonperturbative effect is important: for
$t\overline{t}$ it may be entirely neglected when compared with, e.g.,
the radiative correction.

Eqs. (17), (19) describe rigorously the threshold region, including
radiative corrections, for heavy quarks. The correct treatment, and
inclusion of radiative corrections is essential for {\em open}
$q\overline{q}$ production. This will be discussed in detail in a
forthcoming publication; but an idea may be obtained from Fig. 1 where
we plot various contributions above threshold for $t\overline{t}$
(forgetting, however, the rather important effects of their large width
which could be incorporated with the techniques of refs. 11).

The influence of the threshold region on $\Pi (s)$ with $s$ well below
threshold is an altogether different matter. These effects, that we will
denote by $\Delta_{thr}$, are defined as in QED as the difference
between (17), (19) and the purely perturbative piece and can be obtained
with the help of the dispersion relation (3). We find, integrating up
to\footnote{We have choosen a larger value $v_0$ here because radiative
corrections increase $Im \Pi_{exact}$.} $v_0= \frac{\pi}{\sqrt{2}} C_F
\alpha_s (m^2)$ and neglecting terms of relative order
\[ e^{-2} \sim 10^{-1}, \ \frac{1}{n^3} \log n \meneq 10^{-1}
\mbox{(for $n>2$), } \alpha_s^2 \sim 10^{-2}, \]
and taking the terms dominant (in $v$) above threshold,
\begin{eqnarray}
& {\displaystyle \Delta_{thr} (s) = \frac{N_c}{4\pi^2} \frac{s}{15 m^2}}
\nonumber \\
& {\displaystyle \times \left\{ \frac{15 \pi \zeta (3) C_F^3
\tilde{\alpha}_s (\mu^2)^3}{16}
\left[ 1 + 3 \beta_0 \left( \log \frac{\mu}{mC_F \alpha_s} -\gamma_E
\right) \frac{\alpha_s}{2\pi} \right] \right. } \nonumber \\
& {\displaystyle + \frac{15 \pi^3 C_F^3 \alpha_s (m^2)^2}{16} \left[
\frac{1}{2} \alpha_s
(\mu^2) - \frac{\sqrt{2}}{3} \alpha_s (m^2) + \frac{b}{\pi} \alpha_s
(\mu^2)^2 \right. } \nonumber \\
& {\displaystyle \left. \left. + \frac{\beta_0 \alpha_s (\mu^2)^2}{2\pi}
\left( \log \frac{\mu}{2\pi^2 mC_F \alpha_s} \right) \right] \right\} }
\end{eqnarray}
where
\[ b= \frac{11 C_A - 4T_F n_f}{6} \gamma_E + \frac{31 C_A - 20 T_F
n_f}{36} . \]

The first term in the r.h.s. of (20) comes from the poles, the second
from the region above threshold. To compare with the perturbative
evaluation, eq. (16), we write
\begin{eqnarray}
& {\displaystyle \Pi = \Pi^{(1+2+3)} + \Delta_{thr}} \nonumber \\
& {\displaystyle \equiv \frac{N_c}{4\pi^2} \frac{s}{15m^2} \{ 1 +
\delta_{kin}
+ \delta_2 + \delta_3 + \delta_{NP} + \delta_{pole} + \delta_{thr} \} ,}
\nonumber
\end{eqnarray}
and one has, choosing $\mu^2=s=M^2_Z$,
\begin{eqnarray}
\delta_{kin} & = & \frac{3}{28} \frac{s}{m^2} =  4.0 \times 10^{-2},
\nonumber \\
\delta_2 & = & \frac{205}{54} \frac{C_F \alpha_s}{\pi} = (18.53 \pm
1.61) \times 10^{-2}, \nonumber \\
\delta_3 & = & \frac{205}{54} \frac{C_F \beta_0 \alpha_s^2}{4\pi^2} \log
\frac{\mu^2}{m^2} = - 1.43 \times 10^{-2}, \nonumber \\
\delta_{NP} & = & - \frac{2\pi < \alpha_s G^2>}{21m^4} = -2 \times
10^{-11}, \nonumber \\
\delta_{pole} & = & \frac{15\pi \zeta (3) C_F^3 \tilde{\alpha}_s
(\mu^2)^3}{16} \left[ 1 + 3 \beta_o \left( \log \frac{\mu}{mC_F
\alpha_s} - \gamma_E \right) \frac{\alpha_s}{2\pi} \right] \nonumber \\
	& = & (1.81 \times 10^{-2}) [1+0.08] = 1.96 \times 10^{-2},
\nonumber \\
\delta_{thr} & = & \frac{15\pi^3 C_F^3 \alpha_s (m^2)^2}{16} \left\{
\frac{1}{2} \alpha_s (\mu^2) - \frac{\sqrt{2}}{3} \alpha_s (m^2) \right.
\nonumber \\
    & + & \left. \frac{b}{\pi} \alpha_s^2 (\mu^2) + \frac{\beta_0 \alpha_s
(\mu^2)^2}{2\pi} \left( \log \frac{\mu}{2\pi^2 m C_F \alpha_s} \right)
\right\} = 0.54 \times 10^{-2}.
\end{eqnarray}
Of $\delta_{thr}$, almost 40\% of the value is due to the radiative
correction. The error in $\delta_2$ is due to the error $\alpha_s
(M_Z^2)= 0.115 \pm 0.010$.

It is clear that, just as in the QED case, the effects of the regions
around threshold ($\delta_{pole}$, $\delta_{thr}$) are not only
nominally of higher order than $\delta_2$, $\delta_3$; but numerically
very small compared with the first and of the order of either the {\em
error} in $\delta_2$ or the only partially known therm $\delta_3$. (21)
also shows that, while nonperturbative corrections are negligible, {\em
radiative} ones are quite large (relatively speaking).

We may conclude i) that the threshold effects are small. It would thus
seem that the large effects claimed in e.g. ref. 1 are not supported by
a detailed, rigorous calculation, and ii) that a determination of the
QCD corrections to $\Pi$ precise to the percent level does require
completion of the three loop, $0(\alpha_s^2)$ calculation, improvement
on the determination of $\alpha_s (M^2_Z)$ in addition (and at the same
time as) inclussion of the threshold effects.

\subsection*{Acknowledgements}

The author is very grateful to Profs. R. Akhoury and M. Veltman for very
helpful discussions. Thanks are also due to the hospitality of the
University of Michigan, Randall Lab. of Physics, where this work was
done.

\newpage

\subsection*{Figure Caption}

Fig. 1. Plot of $Im\Pi \times {\mbox constant}$, the constant such that
$Im\Pi \times {\mbox constant} \rightarrow N_c Q_t^2 = 4/3$ for $s
\rightarrow \infty$.

Dotted lines: 1- and 1+2- loop perturbative calculation.
Continuous line: with Fermi function ($F$) and including
radiative
corrections ($F+$ rad). Shaded area: region of integration to get
$\Delta_{thr}$.

\end{document}